\documentstyle[aps]{revtex}

\begin{document}
\title{Evolution of a coherent array of Bose-Einstein Condensates in a magnetic trap}
\author{Hongwei Xiong,$^{1,2}$ Shujuan Liu,$^{1}$ Guoxiang Huang,$^{3,4}$ Zhijun Xu$^{1}$}
\address{$^{1}$Department of Applied Physics, Zhejiang University of Technology, Hangzhou, 310032, China}
\address{$^{2}$Zhijiang College, Zhejiang University of Technology, Hangzhou, 310012, China }
\address{$^{3}$Department of Physics, East China Normal University, Shanghai, 200062, China}
\address{$^{4}$Laboratorie de Physique Th\`{e}orique de la Mati\`{e}re Condens\`{e}e, case 7020, 2 Place Jussieu, 75251\\
Paris Cedex 05, France}
\date{\today}
\maketitle

\begin{abstract}
{\it We investigate the evolution process of the interference pattern for a
coherent array of Bose-Einstein condensates in a magnetic trap after the
optical lattices are switched off. It is shown that there is a decay and
revival of the density oscillation for the condensates confined in the
magnetic trap. We find that, due to the confinement of the magnetic trap,
the interference effect is much stronger than that of the experiment induced
by Pedri et al. (Phys. Rev. Lett, {\bf 87}, 220401), where the magnetic trap
is switched off too. The interaction correction to the interference effect
is also discussed for the density distribution of the central peak.\newline
hongweixiong@hotmail.com\newline
PACS number(s): 03.75.Fi, 05.30.Jp}
\end{abstract}


\section{Introduction}


The development of the technologies of the laser trapping and evaporative
cooling has yielded intriguing Bose-Einstein condensates (BECs) \cite{ALK},
a state of matter in which many atoms are in the same quantum mechanical
state. The realization of BECs has made remarkable theoretical and
experimental advances on this exotic quantum system \cite{Nature,RMP}.
Recently, optical lattices are used to investigate further the unique character of the ultra-cold atoms \cite{SEVERAL}.
The application of the optical lattices to the ultra-cold atoms is very promising, such as
the quantum computing scheme proposed in \cite{JAK,BRE}. Experimentally, the
macroscopic quantum interference effect \cite{ANDER} and thermodynamic
properties \cite{THERMO} of the BECs in the optical lattices have been
investigated thoroughly. In addition, the superfluid and dissipative
dynamics \cite{SUPER,CATA} of a BEC in the optical lattices are
investigated in the experiments too. In particular, through the application
of the optical lattices, the quantum phase transition from a superfluid to a
Mott insulator in Bose-Einstein condensed gases is observed in a recent
experiment \cite{GREL}.

Recently, the expansion of a coherent array of BECs is carried out to
illustrate the interference effect in the experiment by Pedri {\it et al.} 
\cite{PEDRI}, when both the magnetic trap and optical lattices are switched
off. In another experiment by Morsch {\it et al. }\cite{MORSCH}, the
expansion of the condensates is investigated when only the magnetic trap is
switched off. In this paper, we give a theoretical
investigation on the expansion of the condensates when only the optical
lattices are switched off. Due to the confinement of the harmonic potential,
we find that there are several interested phenomena not encountered in the
experiment by Pedri {\it et al. }\cite{PEDRI} and Morsch {\it et al. }\cite
{MORSCH}. In the presence of the harmonic potential, the interference effect
would be much stronger than the case when the magnetic trap is switched off
too. In this situation, researches show that the interaction between atoms would
give important correction, in contrast to the case investigated by Pedri 
{\it et al. }\cite{PEDRI}.


\section{Wave function in harmonic potential after optical lattices are
switched off}


In the experiment conducted by Pedri {\it et al.} \cite{PEDRI}, the external
potential of the Bose gas is given by \cite{PEDRI}:

\begin{equation}
V=\frac{1}{2}m\left( \omega _{x}^{2}x^{2}+\omega _{\bot }^{2}\left(
y^{2}+z^{2}\right) \right) +sE_{R}\cos ^{2}\left( \frac{2\pi x}{\lambda }+%
\frac{\pi }{2}\right) ,  \label{potential}
\end{equation}
The last term represents the external potential due to the presence of the
optical lattices. In the above expression, $\omega _{x}$ and $\omega _{\perp
}$ are the axial and radial frequencies of the harmonic potential,
respectively. In addition, $\lambda $ is the wavelength of the
retroreflected laser beam, and $sE_{R}$ denotes the depth of the optical lattices. For the
optical lattices created by the retroreflected laser beam, the last term has
a period $d=\lambda /2$. In other words, $d$ can be regarded as the distance
between two neighboring wells induced by the optical lattices. In this paper, the experimental parameters
in \cite{PEDRI} are used to calculate various physical quantities, thus it would be useful to give them here.
The experimental parameters in \cite{PEDRI} are $\omega _{x}=2\pi \times 9$ $\rm Hz$,
$\omega _{\perp }=2\pi \times 92$ $\rm Hz$, $\lambda =795$ ${\rm nm}$, 
$E_{R}=2\pi \hbar \times 3.6$ $\rm kHz$, and $s=5$.

Due to the presence of the optical lattices, there are an array of
condensates formed in the combined potential, when the temperature is lower
than the critical temperature. In this work, we shall investigate the case
of the strong tunneling between neighboring BECs, which holds in the
experiment by Pedri {\it et al.} \cite{PEDRI}. In this situation, all the
condensates are fully coherent, and can be described by a single order
parameter. To emphasis the role of the optical lattices, our researches are
carried out mainly on the character of the coherent array of condensates in
the $x-$direction.

In the presence of a magnetic trap, the number of atoms in each well
should be different. Based on the analysis of the 3D model
of the condensates in the combined potential \cite{PEDRI}, the ratio between
the number of condensed atoms in $k-$th and central wells is given by $%
N_{k}/N_{0}=\left( 1-k^{2}/k_{M}^{2}\right) ^{2}$, where $2k_{M}+1$ \cite
{PEDRI} represents the total number of condensates induced by the optical
lattices. In this situation, using the Gaussian approximation in the $x-$%
direction for each well, the normalized wave function in coordinate space
takes the form

\begin{equation}
\varphi _{0}\left( x\right) =A_{n}\sum_{k=-k_{M}}^{k_{M}}\left( 1-\frac{k^{2}%
}{k_{M}^{2}}\right) \exp \lbrack -\left( x-kd\right) ^{2}/2\sigma
^{2}\rbrack ,  \label{non-dis}
\end{equation}
where $\sigma $ denotes the width of the condensate in each well. It can be
calculated by numerical minimization of the energy of the condensates \cite
{PEDRI}. In the following calculations in this paper, $\sigma=0.25d$ \cite{PEDRI} is used for $s=5$.
In the above expression, the normalized constant $A_{n}$ takes the form

\begin{equation}
A_{n}=\frac{1}{\sqrt{\left( 16k_{M}^{4}-1\right) /15k_{M}^{3}}\pi
^{1/4}\sigma ^{1/2}}.  \label{non-norm}
\end{equation}

It is well known that once the wave function at an initial time is
known, the wave function at a later time can be obtained through the
following integral equation \cite{FEYN}:

\begin{equation}
\varphi _{0}\left( x,t\right) =\int_{-\infty }^{\infty }K(x,t;y,t=0)\varphi
_{0}\left( y,t=0\right) dy,  \label{evolution}
\end{equation}
where $\varphi _{0}\left( y,t=0\right) $ is the wave function at the initial
time $t=0$ which is given by Eq. (\ref{non-dis}), and $K(x,t;y,t=0)$ is the
well known propagator. For the atoms in the harmonic potential, the
propagator can be expressed as \cite{FEYN}:

\begin{equation}
K(x,t;y,t=0)=\left[ \frac{m\omega _{x}}{2\pi i\hbar \sin \omega _{x}t}%
\right] ^{1/2}\exp \left\{ \frac{im\omega _{x}}{2\hbar \sin \omega _{x}t}%
\left[ \left( x^{2}+y^{2}\right) \cos \omega _{x}t-2xy\right] \right\} .
\label{propagator-harmonic}
\end{equation}
From the formulas (\ref{non-dis}), (\ref{evolution}) and (\ref
{propagator-harmonic}), after a straightforward calculation, one obtains the
analytical result of the wave function confined in the magnetic trap

\[
{\ \varphi }_{0}{\left( x,t\right) =A_{n}\sqrt{\frac{1}{\sin \omega
_{x}t\left( {\rm ctg}\omega _{x}t+i\gamma \right) }}\sum_{k=-k_{M}}^{k_{M}}%
\left( 1-\frac{k^{2}}{k_{M}^{2}}\right) \exp \left[ -\frac{\left( kd\cos
\omega _{x}t-x\right) ^{2}}{2\sigma ^{2}\sin ^{2}\omega _{x}t\left( {\rm ctg}%
^{2}\omega _{x}t+\gamma ^{2}\right) }\right] \times } 
\]

\begin{equation}
\exp \left[ -\frac{i\left( kd\cos \omega _{x}t-x\right) ^{2}{\rm ctg}\omega
_{x}t}{2\gamma \sigma ^{2}\sin ^{2}\omega _{x}t\left( {\rm ctg}^{2}\omega
_{x}t+\gamma ^{2}\right) }\right] \exp \left[ \frac{i\left( x^{2}\cos \omega
_{x}t+k^{2}d^{2}\cos \omega _{x}t-2xkd\right) }{2\gamma \sigma ^{2}\sin
\omega _{x}t}\right] ,  \label{evolution2}
\end{equation}
where we have introduced a dimensionless parameter $\gamma =\hbar /m\omega
_{x}\sigma ^{2}$. Assuming $N$ denotes the total number of particles in the
condensates, the density distribution in $x-$direction is $n\left(
x,t\right)=N\left| {\varphi }_{0}{\left( x,t\right) }\right| ^{2}$.

\vspace{1pt}

\section{Periodicity of the Density Distribution and the Motion of {$n=\pm 1$%
} Peak}


Due to the confinement of the harmonic potential, the density distribution
in $x-$direction should exhibit a periodic character. From Eq. (\ref
{evolution2}), it is easy to find that the period of the density
distribution $n\left( x,t\right) $ is determined by $\omega _{x}T=\pi $. For
the experiment in \cite{PEDRI}, $\omega _{x}=2\pi \times 9$ ${\rm Hz}$, this
means that the period of the density distribution is given by $500/9$ ${\rm %
ms}$.

In Fig. 1, displayed is $n\left( x=0,t\right) $ when only
the optical lattices are switched off. The periodicity of the density is
clearly shown in the figure and in agreement with the analytical result
given by $T=\pi /\omega _{x}$. We see that the density at $x=0$ reaches a
maximum value at time $t_{m}=\left( 2m-1\right) \pi /2\omega _{x}$, with $m$
positive integer. At time $t_{m}$, the wave function takes the form

\begin{equation}
\varphi _{0}\left( x,t_{m}\right) =A_{n}\sqrt{\frac{1}{i\gamma }}%
\sum_{k=-k_{M}}^{k_{M}}\left( 1-\frac{k^{2}}{k_{M}^{2}}\right) \exp \left[ -%
\frac{x^{2}}{2\sigma ^{2}\gamma ^{2}}-i\frac{xkd}{\gamma \sigma ^{2}}\right].
\label{wavefunctionfun}
\end{equation}
The maximum density at $x=0$ is then

\begin{equation}
n\left( x=0,t_{m}\right) =\frac{NA_{n}^{2}}{\gamma }\left[
\sum_{k=-k_{M}}^{k_{M}}\left( 1-\frac{k^{2}}{k_{M}^{2}}\right) \right] ^{2}.
\label{maximum}
\end{equation}
In the case of $k_{M}>>1$, the above expression can be approximated as:

\begin{equation}
n\left( x=0,t_{m}\right) \approx N\alpha _{x-ideal}^{2},  \label{maximum1}
\end{equation}
where $\alpha _{x-ideal}^{2}$ is given by

\begin{equation}
\alpha _{x-ideal}^{2}=\frac{5k_{M}m\omega _{x}\sigma }{3\pi ^{1/2}\hbar }.
\label{alphaideal}
\end{equation}

In the experiments, the depth of the optical lattices can be changed through the
variation of the parameter $s$. From \cite{PEDRI}, we know 
$\sigma \propto 1/s^{1/4}$ and $k_{M}\propto 1/\sigma ^{1/5}$. Therefore,
$n\left( x=0,t_{m}\right) \propto 1/s^{1/5}$. This shows that in the non-interacting model,
the maximum density at $x=0$ decreases with the increase of the depth of the optical lattices.

In Figures 2 (a)-(d), we show the evolution of the density distribution of
the condensates confined in the magnetic trap, after the optical lattices
are switched off. The density distributions are shown at $t=0$, $0.1\pi
/\omega _{x}$, $0.3\pi /\omega _{x}$, and $0.5\pi /\omega _{x}$ in these
figures. The motion of the $n=\pm 1$ peaks is clearly shown in these
figures. The oscillation motion of the $n=\pm 1$ peaks is due to the
confinement of the harmonic potential. In fact, the motion of the $n=\pm 1$
peaks can be described very well using the classical harmonic motion. Using
the classical harmonic motion, the motion of the $n=\pm 1$ peaks is
determined by the following expression:

\begin{equation}
x_{n=\pm 1}\left( t\right) =\pm \frac{2\pi \hbar }{m\omega _{x}d}\cos \left(
\omega _{x}t-\frac{\pi }{2}\right) .  \label{harmonic motion}
\end{equation}
When obtaining the above formula, we have used the fact that the momentum
distribution is characterized by sharp peaks at the values $p_{x}=n2\pi
\hbar /d$ \cite{PEDRI}. The solid line in Fig. 3 shows the harmonic motion of the $n=1$
peak using the above formula, while the circles show the result 
given by Eq. (\ref{evolution2}). We see that the classical
harmonic motion agrees quite well with the result given by Eq. (\ref{evolution2}). In a sense, Eq. (\ref{harmonic motion})
describes the motion of the center of mass of the $n=\pm 1$ peak. Thus, we anticipate that the interaction
between atoms will not change the motion of the $n=\pm 1$, although it will affect the density and width of the 
$n=\pm 1$ peak.

From Fig. 2(d) we see that at time $t_{m}$ the density distribution in $x-$direction 
exhibits a very sharp peak at the center of the magnetic trap. The
maximum density of the central peak is given by Eq. (\ref{maximum1}). As a
comparison, assume there are $N$ atoms confined in an identical magnetic
trap, but there are no optical lattices to induce the interference effect.
In this situation, in $x-$direction, the density distribution at $x=0$ is
given by

\begin{equation}
n_{mag}\left( x=0\right)=\frac{\pi \mu_{mag}^{2}}{gm\omega _{\perp }^{2}},
\label{second}
\end{equation}
where $\mu_{mag} $ is the chemical potential of the Bose gas confined in the
magnetic trap. The ratio between $n\left( x=0,t_{m}\right) $ and $%
n_{mag}\left( x=0\right)$ is then

\begin{equation}
\frac{n\left( x=0,t_{m}\right) }{n_{mag}\left( x=0\right)}=\frac{Ngm\omega
_{\perp }^{2}\alpha _{x-ideal}^{2}}{\pi \mu_{mag}^{2}}.  \label{ratio1}
\end{equation}
For the experimental parameters in \cite{PEDRI}, $n\left( x=0,t_{m}\right)
/n_{mag}\left( x=0\right)=29.6$. This shows clearly that there is a very
strong interference effect for the case considered here. 

If both the
magnetic trap and optical lattices are switched off, there is also a sort of
interference effect. The maximum value of the density distribution in this
case is given by

\begin{equation}
n_{bs}\left( x=0\right)=\frac{N_{0}}{\sqrt{\pi }\sigma }.  \label{third}
\end{equation}
From Eq. (\ref{third}), $n_{bs}\left( x=0\right) \propto s^{1/4}$. In contrast to the case where
only the optical lattices are switched off, $n_{bs}\left( x=0\right) $ decreases with the decrease
of the depth of the optical lattices.

The ratio between $n_{bs}\left( x=0\right)$ and $n_{mag}\left( x=0\right)$
is then

\begin{equation}
\frac{n_{bs}\left( x=0\right) }{n_{mag}\left( x=0\right) }=\frac{%
N_{0}gm\omega _{\perp }^{2}}{\pi ^{3/2}\sigma \mu _{mag}^{2}}.
\label{ratio2}
\end{equation}
For the experimental parameters in \cite{PEDRI}, we have $n_{bs}\left(
x=0\right) /n_{mag}\left( x=0\right) =2.2$. From Eqs. (\ref{maximum1}) and 
(\ref{third}), $n\left( x=0,t_{m}\right) /n_{bs}\left( x=0\right) =13.7$.
Therefore, when only the optical lattices are switched off, the interference
effect would be much stronger than the case when the magnetic trap is
switched off too. When only the optical lattices are switched off, at $t_{m}$
the density of the central peak is very high, and we anticipate that in this
situation the interaction between atoms would give important correction.


\section{Decay and revival of the density oscillation}


\vspace{1pt}From Fig. 1, we see that there is a phenomenon of decay and
revival of the density oscillation at $x=0$. We now turn to discuss this
unique character. To proceed, it is useful to introduce an important time
scale which determines when the interference between two neighboring
condensates begins to occur. Before the magnetic trap and optical lattices
are switched off, from the Gaussian approximation of the condensates in each
well, the width of the condensates in each well is given by

\begin{equation}
\Delta x_{0}^{2}=\frac{\int x^{2}\exp \left[ -x^{2}/\sigma ^{2}\right] dx}{%
\int \exp \left[ -x^{2}/\sigma ^{2}\right] dx}.  \label{width}
\end{equation}
It is easy to get $\Delta x_{0}=\sigma /\sqrt{2}$ from the above formula.
When the optical lattices are switched off, for time much smaller than $\pi
/\omega _{x}$, the condensates can be approximated as a free expansion and
the width of the condensates would increase in this situation. Based on the
analysis of the spreading of the wave packet, the width of each condensate
is given by

\begin{equation}
\Delta x\left( t\right) =\Delta x_{0}\sqrt{1+\frac{\hbar ^{2}t^{2}}{%
m^{2}\Delta x_{0}^{4}}}.  \label{width-t}
\end{equation}
When $\Delta x\left( t\right) =d$, the condensates in neighboring wells
begin to interfere with each other. By setting $\Delta x\left( t\right) =d$
in the above formula, we obtains a time scale $t_{w}$ which determines when
the interference between neighboring condensates begins to occur. From Eq. (%
\ref{width-t}), it is easy to find that the following analytical result of $%
t_{w}$ can give a rather well approximation

\begin{equation}
t_{w}=\frac{\sigma dm}{\sqrt{2}\hbar }.  \label{tw}
\end{equation}
From Eqs. (\ref{width-t}) and (\ref{tw}), for $t>t_{w}$, one gets the
following useful result:

\begin{equation}
\Delta x\left( t\right) =\frac{t}{t_{w}}d.  \label{width-t-c}
\end{equation}

As illustrated in Fig. 1, the oscillation of the density at $x=0$ will cease
ultimately when $t>k_{M}t_{w}$. However, when the time approaches $\pi
/\omega _{x}$, the density oscillation will reappear. Note that the time for
the revival of the density oscillation is determined solely by the axial
frequency of the harmonic potential. This shows that the confinement of the
harmonic potential plays a crucial role for the revival phenomenon of the
density oscillation. To verify further the decay and revival of the density
oscillation, Fig. 4 shows the density at $x=k_{M}d/2$. Analogous decay and
revival of the density oscillation are illustrated clearly in the figure.
However, the oscillation of the density disappears at a longer time $%
t=1.47k_{M}t_{w}$, in comparison with the case at $x=0$.

In Fig. 5, we display the time of the disappearance of the density
oscillation for different locations in the region $0<t<\pi /\omega _{x}$. We
can give a rather simple interpretation for this result. When the
optical lattices are switched off, the width of the expanding condensates in
each well will increase. For the location at $x=0$, there are more and more
expanding BECs interfere at this point with the development of the time.
This is the reason why the density at the point $x=0$ will oscillate
intensely. When $t>k_{M}t_{w}$, however, all expanding BECs have
participated in the interference at the point $x=0$. Therefore, the
oscillation of the density at $x=0$ will cease at time longer than $%
k_{M}t_{w}$. Generalizing this result, the time for the disappearance of the
density oscillation for different locations is given by the following simple
expression:

\begin{equation}
t=(k_{M}+x/d)t_{w}.  \label{time-dis}
\end{equation}
The solid line in Fig. 5 displays the above analytical result. We see from
Fig. 5 that this simple expression agrees well with the result given by Eq. (\ref{evolution2}).
Maybe the slight difference from the result given by Eq. (\ref{evolution2}) lies in the fact that
we do not account for the effect of non-uniform atom distribution in each
well when obtaining Eq. (\ref{time-dis}).

It is worth pointing out that the disappearance of the density oscillation
do not mean the disappearance of the interference effect between the
expanding condensates. When $0<t<<\left( 2k_{M}+1\right) t_{w}$, there are
only several expanding condensates interfering with each other, and there
are interference fringes (or density oscillation) in this situation. For $%
\left( 2k_{M}+1\right) t_{w}<t<\pi /\omega _{x}-\left( 2k_{M}+1\right) t_{w}$,
however, all expanding condensates will interfere with each other, and
this means the emergence of the diffraction fringes. In fact, $n=0$ and $%
n=\pm 1$ peaks in Fig. 2 (a)-(d) should be regarded as the diffraction
fringes, rather than the interference fringes. Note that the phenomena of
diffraction and interference are basically equivalent. Different from the
interference phenomenon, however, the diffraction phenomenon should be
regarded as a consequence of interference from many coherent wave sources.
In a sense, Eq. (\ref{time-dis}) gives the time for the emergence of the
diffraction fringes, which means the disappearance of the density
oscillation.


\section{Interaction correction to the central peak at $t_{m}$}


In the case of non-interacting model, we have shown that, at time $t_{m}$,
there would be a sharp central peak in the magnetic trap. In this case, the
interaction between atoms can not be simply omitted, in contrast to the case
when the magnetic trap is switched off too. At time $t_{m}$, the central
peak of the density in $x-$direction can be approximated as a Gaussian
distribution. After the optical lattices are switched off, using
Thomas-Fermi approximation in the radial direction, the square of the
modulus of the 3D wave function at time $t_{m}$ takes the form

\begin{equation}
\left| \varphi _{0}\left( x,r_{\perp },t_{m}\right) \right| ^{2}=\alpha
_{x}^{2}\alpha _{\perp }^{2}\exp \left[ -\frac{2x^{2}}{R_{x}^{2}}\right]
\left( 1-\frac{r_{\perp }^{2}}{R_{\perp }^{2}}\right),  \label{3Dwave}
\end{equation}
where $R_{\perp }^{2}=\sqrt{2\mu _{0}/m\omega _{\perp }^{2}}$ with $\mu
_{0}=m\omega _{x}^{2}k_{M}^{2}d^{2}/2$ \cite{PEDRI}. In the above
expression, $\alpha _{x}=(2/\pi )^{1/4}/\sqrt{R_{x}}$ and $\alpha _{\perp }=%
\sqrt{2/\pi R_{\perp }^{2}}$ are the normalized constants in the axial and
radial directions, respectively. Obviously, $N\alpha _{x}^{2}$
represents the density $n\left( x=0,t_{m}\right) $ in the $x-$direction. Due
to the repulsive interaction between atoms, we anticipate that $\alpha
_{x}^{2}<\alpha _{x-ideal}^{2}$.

Assume $E_{int}$ and $E_{kin}$ are the interaction energy and kinetic energy
of the central peak at $t_{m}$, respectively. The interaction energy of the
central peak is given by

\begin{equation}
E_{int}=\frac{gN^{2}}{2}\int \left| \varphi _{0}\left( x,r_{\perp
},t_{m}\right) \right| ^{4}dV=\frac{\sqrt{2}gN^{2}\alpha _{x}^{2}}{3\pi
R_{\perp }^{2}}\text{.}  \label{intenergy}
\end{equation}
Assuming the total energy of the condensates is $E_{all}$, we have

\begin{equation}
E_{kin}+E_{int}+E_{ho}=E_{all},  \label{totalenergy}
\end{equation}
where $E_{ho}$ is the potential energy of the condensates. For the central
peak, $E_{ho}$ can be omitted safely. In the case of non-interacting model, $%
E_{all}$ will transform fully to the kinetic energy and potential energy of
the condensates once the optical lattices are switched off. Due to the
presence of the repulsive interaction between atoms, the maximum density of
the central peak would be smaller than the result of the non-interacting
model.

Note that the kinetic energy of the central peak can not be calculated through $%
N\int \frac{\hbar ^{2}}{2m}\left( \nabla \sqrt{\left| \varphi _{0}\left(
x,r_{\perp },t_{m}\right) \right| ^{2}}\right) ^{2}dV$, because the phase
factor is different for different well (See Eq. (\ref{wavefunctionfun})).
From the uncertainty relation, assume $E_{kin}\propto 1/R_{x}^{2}\propto
\alpha _{x}^{4}$. In the presence of repulsive interaction, we have

\begin{equation}
E_{kin}=\frac{\alpha _{x}^{4}}{\alpha _{x-ideal}^{4}}E_{all}.
\label{relation1}
\end{equation}
From Eqs. (\ref{intenergy}), (\ref{totalenergy}) and (\ref{relation1}), one obtains the
following equation to determine $\alpha _{x}^{2}$:

\begin{equation}
\beta ^{2}+\theta \beta -1=0,  \label{equationbeta}
\end{equation}
where $\beta =(\alpha _{x}/\alpha _{x-ideal})^{2}$. The value of $\beta $
reflects how the repulsive interaction between atoms reduces the density of
the central peak. In the above expression, the dimensionless parameter $%
\theta =E_{int}\left( \alpha _{x}=\alpha _{x-ideal}\right) /E_{all}$. From
Eq. (\ref{equationbeta}), we have

\begin{equation}
\beta =\frac{-\theta +\sqrt{\theta ^{2}+4}}{2}.  \label{beta}
\end{equation}

Now let us turn to discuss the total energy of the condensates which is
necessary to calculate the value of $\beta $. The total energy of the
condensates can be obtained through the sum of the energy of the condensates
in each well before the optical lattices are switched off. Before the
optical lattices are switched off, the normalized wave function in $k-$th
well takes the form

\begin{equation}
\varphi _{0k}\left( x,r_{\perp }\right) =\varphi _{0k}\left( x\right)
\varphi _{0k}\left( r_{\perp }\right)  \label{kwave}
\end{equation}
where

\begin{equation}
\varphi _{0k}\left( x\right) =\left( \frac{1}{\sqrt{\pi }\sigma }\right)
^{1/2}\exp \left[ -\frac{\left( x-kd\right) ^{2}}{2\sigma ^{2}}\right] ,
\label{kwavex}
\end{equation}
and $\varphi _{0k}\left( r_{\perp }\right) $ is the normalized wave function
in the radial direction. From Eq. (\ref{kwave}), we have

\[
{\ {E_{all}}=\sum_{k=-k_{M}}^{k_{M}}\left\{ N_{k}\int \varphi _{0k}\left(
x,r_{\perp }\right) \left[ -\frac{\hbar ^{2}}{2m}\nabla ^{2}+\frac{1}{2}%
m\omega _{xe}^{2}x^{2}+\frac{1}{2}m\omega _{\perp }^{2}r_{\perp }^{2}\right]
\varphi _{0k}\left( x,r_{\perp }\right) dV\right. }
\]

\begin{equation}
\left. +\frac{gN_{k}^{2}}{2}\int \left| \varphi _{0k}\left( x,r_{\perp
}\right) \right| ^{4}dV\right\} ,  \label{totalener}
\end{equation}
where $\omega _{xe}$ is the effective harmonic frequency in the $x-$%
direction of the well induced by the optical lattices. The last term in the
above expression represents the interaction energy of the condensates in
each well, and it is easy to verify that it can be omitted safely. In
addition, the kinetic energy and potential energy in the radial direction
can be also omitted because $\omega _{xe}>>\omega _{\perp }$, and $\hbar
\omega _{xe}>>\mu _{0}$. In this case, one gets

\begin{equation}
E_{all}\approx N\left( \frac{\hbar ^{2}}{4m\sigma ^{2}}+\frac{1}{4}m\omega
_{xe}^{2}\sigma ^{2}\right) .  \label{a-total-ener}
\end{equation}

For the experimental parameters in \cite{PEDRI}, from the formulas (\ref
{intenergy}), (\ref{beta}), and (\ref{a-total-ener}), the 
calculation shows that $\beta =0.80$. This shows clearly that the repulsive
interaction between atoms would reduce the density of the central peak at $%
t_{m}$.

With the variation of the parameter $s$,, the maximum density of the central peak
can be also calculated based on the method given here. In the interacting model, the solid line
in Fig. 6 shows the ratio between
$n\left( x=0,t_{m},s\right) $ and $n\left( x=0,t_{m},s=5\right) $. The result of the non-interacting model is also shown
in the figure.

\section{Discussion and Conclusion}


In brief, the evolution process of the condensates is investigated after the
optical lattices are switched off. We find that the density oscillation
exhibits a phenomenon of decay and revival, based on the numerical result of
the evolution of the density distribution. The decay of the density
oscillation is interpreted as the emergence of the diffraction phenomenon,
which is regarded as a consequence of interference from a lot of coherent
expanding condensates. Due to the confinement of the harmonic potential,
there is a periodic character of the density distribution, and it is this
periodic character which leads to the revival of the density oscillation. In
contrast to the condensates in the magnetic trap, there is no revival of the
density oscillation, when both the magnetic and optical lattices are
switched off. In addition, in the case of non-interacting model, it is shown
that the maximum value of the density distribution at $x=0$ would be
approximately $30$ times larger than the case when there are no optical
lattices to induce the interference effect. 

It is shown here that the repulsive interaction between atoms has an effect
of reducing the maximum density of the central peak. In a real experiment in future,
maybe the experimental result of the maximum density of the central peak
would be smaller than the theoretical prediction given here, because it is
possible that there is a loss of the total energy of the condensates during
the process of the removal of the optical lattices. For the attractive
interaction such as $Li$, the role of interaction would become very
important when the atoms are confined by the combined potential. For
example, when only the optical lattices are removed, based on the
non-interacting model, the density at $x=0$ would increase largely due to
the interference and confinement of the magnetic trap. In addition, due to
the attractive interaction between atoms, the density of the central peak
would increase rapidly. In this situation, it is possible that the
condensates would collapse and even explode in a subsequent time, in
analogous with the dynamic process of the collapsing and exploding atoms 
\cite{EXPL} by switching the interaction from repulsive to attractive.


\section*{Acknowledgments}


This work was supported by the Science Foundation of Zhijiang College,
Zhejiang University of Technology, and Natural Science Foundation of
Zhejiang Province. One of us (G. H.) is indebted to National Natural Science
Foundation of China under Grant No. 19975019, and the French Ministry of
Research for a visiting grant at Universit\`{e} Paris 7.

\newpage

\section*{Figure Caption}

Fig. 1 Displayed is the density of the condensate at $x=0$ vs time $t$,
after the optical lattice is switched off. Here the density $n(x=0,t)$ is in
units of $NA_{n}^{2}$. We can see clearly in the figure that there is a
decay and revival of the density oscillation. In addition, there is a
periodicity of the density due to the confinement of the magnetic trap.

Fig. 2 (a)-(d) show the evolution of the density distribution with time $t$,
after the optical lattices are switched off. The density distributions are
shown at $t=0$, $0.1\pi /\omega _{x}$, $0.3\pi /\omega _{x}$, and $0.5\pi
/\omega _{x}$. The emergence and motion of the $n=\pm 1$ peaks are clearly
shown in these figures. Here the density distribution $n(x,t)$ is in units
of $NA_{n}^{2}$, while the location $x$ is in units of $d$, {\it i.e.} the
distance between two neighboring condensates.

Fig. 3 Displayed is the motion of the $n=1$ peak, after the optical lattices
are switched off. Here the location $x\left( t\right) _{n=1}$ of $n=1$ peak
is in units of $d$. The solid line is the result calculated from the
classical harmonic motion given by Eq. (\ref{harmonic motion}). The squares
show the motion of $n=1$ peak obtained from the numerical result given by
Eq. (\ref{evolution2}). We see that the classical harmonic motion agrees
quite well with the numerical result.

Fig. 4 Displayed is the density of the condensate at $x=k_{M}d/2$ vs time $t$,
after the optical lattices are switched off. Here the density $n(x,t)$ is in
units of $NA_{n}^{2}$. There is a phenomenon of the decay and revival of the
density oscillation.

Fig. 5 Displayed is the time of the disappearance of the density oscillation
for different locations. Here the location $x$ is in units of $d$, while the
time $t$ is in units of the time scale $t_{w}$. The solid line is obtained
from the formula (\ref{time-dis}), while the squares show the result
obtained directly from the numerical result given by Eq. (\ref{evolution2}).
We see that the analytical formula (\ref{time-dis}) can give a well
description for the disappearance of the density oscillation.

Fig. 6 Displayed is the ratio between $n\left( x=0,t_{m},s\right) $ and $n\left( x=0,t_{m},s=5\right) $ for
the interacting and non-interacting models.

\end{document}